\documentclass[aps,prd,reprint,preprintnumbers,superscriptaddress,nofootinbib,floatfix]{revtex4-2}

\usepackage{amsmath,amssymb,amsfonts,bm}
\usepackage{graphicx}
\usepackage{xcolor}
\usepackage{booktabs}
\usepackage[colorlinks=true,linkcolor=blue,citecolor=blue,urlcolor=blue]{hyperref}

\newcommand{\dd}{\mathrm{d}}
\newcommand{\kms}{\,\mathrm{km\,s^{-1}}}

\begin{document}

\title{A Peccei--Quinn Origin for Inelastic Electroweak Dark Matter after LUX-ZEPLIN}

\author{Luca Visinelli}
\email{lvisinelli@unisa.it}
\affiliation{Dipartimento di Fisica ``E.R.\ Caianiello'', Universit\`a degli Studi di Salerno, Via Giovanni Paolo II 132, 84084 Fisciano (SA), Italy}
\affiliation{Istituto Nazionale di Fisica Nucleare, Gruppo Collegato di Salerno, Sezione di Napoli, Via Giovanni Paolo II 132, 84084 Fisciano (SA), Italy}

\date{\today}

\begin{abstract}
The LUX-ZEPLIN (LZ) experiment has reported a nuclear recoil event reconstructed at $E_R=248\pm23_{\rm stat}\pm23_{\rm sys}{\rm\,keV}$ in a dedicated high-recoil search. We investigate an inelastic electroweak (EW) doublet interpretation in which spontaneous Peccei--Quinn (PQ) breaking leaves a residual parity that stabilizes the lighter neutral state and generates its Majorana splitting. A PQ-charged singlet fermion acquires a Majorana mass from the PQ-breaking vacuum expectation value and, once integrated out, induces the dimension-five operator that splits a vectorlike EW doublet, while a minimal KSVZ sector supplies the QCD anomaly. For $v_S=4.4\times10^{11}{\rm\,GeV}$ and $M_N\simeq50{\rm\,TeV}$, the LZ-motivated splitting requires $y\simeq0.023$. In a standard thermal history, the EW doublet is subdominant. Three kinematic benchmarks with $M_D=297.0$, $396.4$, and $549.7{\rm\,GeV}$, $\delta_\chi\simeq304$--$343{\rm\,keV}$, and $\xi_\chi\simeq0.073$--$0.250$ each yield about one recoil event, while their accepted spectra peak below the observed recoil energy, so they are interpreted as kinematic rather than spectral best-fit points. The remaining abundance can be supplied by the QCD axion through vacuum misalignment. We recompute solar capture for the subdominant thermal WIMP component and compare the resulting annihilation signal with the IceCube $W^+W^-$ limit, finding the LZ-motivated standard-thermal branch excluded under the assumptions adopted here. The excited state has a millisecond-scale radiative lifetime and kilometer-scale decay length, leaving the LZ topology dominated by a single nuclear recoil while allowing complementary luminous signals after terrestrial upscattering.
\end{abstract}

\maketitle

\section{Introduction}
\label{sec:intro}

Weakly interacting massive particles (WIMPs) and the QCD axion are two of the best-developed particle-physics approaches to dark matter (DM). The QCD axion follows from the Peccei--Quinn (PQ) solution of the strong-CP problem~\cite{Peccei:1977hh,Peccei:1977ur,Weinberg:1977ma,Wilczek:1977pj} and is naturally light in invisible-axion realizations~\cite{Kim:1979if,Shifman:1979if,Dine:1981rt,Zhitnitsky:1980tq}. A hypercharged Dirac electroweak (EW) doublet is excluded as ordinary halo DM by its diagonal $Z$ coupling, while a small Majorana splitting renders the neutral current off diagonal and the scattering endothermic~\cite{Tucker-Smith:2001myb,Tucker-Smith:2004mxa}. This pseudo-Dirac mechanism is realized for Higgsinos with heavy gauginos and more generally for hypercharged EW multiplets~\cite{Nagata:2014wma,Barello:2014uda,Bramante:2016rdh}. Endothermic kinematics suppresses the conventional low-energy recoil spectrum and can instead populate the high-recoil region, with the rate controlled by the fastest particles in the local DM distribution.

The LZ Collaboration has performed a dedicated high-recoil search using a $2.84$ tonne-year exposure~\cite{LZ:2026axp}. One event, recorded on 16 June 2023 in a region of low modeled background expectation, has properties consistent with a $248\pm23_{\rm stat}\pm23_{\rm sys}$,keV xenon recoil. Across the effective interactions and inelastic families tested, the maximum local significance is $3.4\sigma$ and the global significance after the look-elsewhere correction is $2.6\sigma$. The event is not evidence for DM, but its high recoil energy and the absence of a corresponding low-energy excess make inelastic scattering with a mass splitting of a few hundred keV a particularly interesting interpretation. A broad range of explanations has subsequently been explored, including endothermic and exothermic scattering, Higgsino and other EW multiplet realizations, absorption mechanisms, singlet--doublet constructions, and non-standard production or acceleration scenarios~\cite{Su:2026rwz,Fan:2026kxx, Freese:2026sga,Wu:2026nhi,Lou:2026idn, DiMauro:2026ldr,Pospelov:2026ewn, Yin:2026jnn,Chattopadhyay:2026ryw, deLima:2026shq,Ahmed:2026qjg, Palmisano:2026kuj,Lee:2026jxl, Jeesun:2026vzo}. These studies have also identified several characteristic tests of an inelastic interpretation. The high-energy LZ sideband can strongly constrain models with a fixed EW strength interaction, the near-threshold regime can generate a pronounced and non-sinusoidal annual modulation, and related EW multiplets occupy correlated trajectories in the mass--splitting plane~\cite{Rodd:2026tyn,McCabe:2026crm,Smirnov:2026aqk}. Solar gravitational acceleration can reopen otherwise suppressed endothermic channels and lead to stringent neutrino-telescope limits~\cite{Pospelov:2026ewn}, while model-independent comparisons of accelerated and endothermic scenarios emphasize the discriminating power of the recoil spectrum itself~\cite{Mahapatra:2026glu}.

Among the particle physics realizations motivated by the LZ event, nearly pure Higgsinos and more general EW multiplets are especially predictive because their off-diagonal $Z$ interaction is fixed by electroweak gauge invariance~\cite{Fan:2026kxx,Freese:2026sga,Wu:2026nhi,Smirnov:2026aqk}. PQ-symmetric high-scale supersymmetric constructions connect this phenomenology to the strong-$CP$ problem and to mixed axion--Higgsino DM~\cite{Yin:2026jnn}, while singlet--doublet realizations can suppress the inelastic transition through mixing~\cite{Lee:2026jxl}. Our construction provides a different connection between PQ breaking and the low-energy pseudo-Dirac structure. Spontaneous PQ breaking leaves a residual parity that stabilizes the lighter EW state and generates the Majorana mass of a heavy neutral singlet. Integrating out this singlet induces the dimension-five operator that splits a vectorlike EW doublet, while a minimal KSVZ colored pair supplies the QCD anomaly~\cite{Kim:1979if,Shifman:1979if}. In contrast with PQ-symmetric Higgsino constructions in which the PQ sector controls supersymmetric Higgs mass parameters, the PQ-breaking scale enters here directly in the origin of the Majorana splitting. The two light states remain almost pure EW doublets, so the inelastic $Z$ interaction is fixed rather than adjusted to reproduce the LZ rate. Related multicomponent axion--WIMP scenarios include mixed axion--neutralino cosmologies~\cite{Baer:2011hx,Bae:2013hma} and non-supersymmetric models in which residual PQ symmetries stabilize a massive DM component~\cite{Dasgupta:2013cwa,Longas:2023bvq}.

This fixed interaction makes the interplay between direct detection, cosmology, and complementary constraints particularly restrictive. In a standard thermal history, a sub-TeV EW doublet constitutes only a fraction of the observed DM abundance, while the QCD axion already present in the KSVZ completion can supply the complementary component through vacuum misalignment. We identify an approximately continuous standard thermal branch on which the reduced EW relic abundance is compensated by the fixed weak interaction and near-threshold endothermic kinematics, yielding an order-one LZ event rate. Three kinematic reference benchmarks with $M_D=297.0$, $396.4$, and $549.7$,GeV and splittings in the range $\delta_\chi\simeq304$--$343$,keV illustrate this regime, for which standard EW freeze-out supplies approximately 7--25\% of the DM density. For a representative value $v_S\simeq4.4\times10^{11}$,GeV, the remaining abundance can be supplied by the QCD axion for an initial misalignment angle of order unity. The accepted recoil spectra are evaluated independently of the kinematic construction and peak below the reference values $E_R^\star=248-\sigma_E$, $248$, and $248+\sigma_E$,keV, with $\sigma_E\simeq32.5$,keV, so these points should be regarded as kinematic benchmarks rather than spectral best fits. The standard thermal branch extends continuously toward the $\sim1.1$,TeV mass at which the EW doublet can account for essentially all of the DM, with increasing $M_D$ accompanied by a larger inelastic splitting~\cite{Cirelli:2005uq,Nagata:2014wma}.

We also examine the radiative decay of the excited state and complementary signatures from terrestrial upscattering~\cite{Eby:2019mgs,Graham:2024syw} and collider production~\cite{Fukuda:2017jmk,Fukuda:2019kbp}. Most importantly, we recompute solar capture for the subdominant thermal EW component, including the same halo assumptions and finite momentum nuclear response used in the terrestrial analysis, and compare the resulting annihilation signal with the existing IceCube $W^+W^-$ limit~\cite{Catena:2018vzc,Pospelov:2026ewn}. Under the standard halo, thermal abundance, and annihilation assumptions adopted here, the LZ-motivated standard thermal branch is excluded by the solar neutrino constraint. The resulting picture links the microscopic origin of the pseudo-Dirac splitting to PQ breaking, mixed axion--WIMP cosmology, the high-recoil LZ phenomenology, and complementary terrestrial and solar probes within a single construction.

\section{PQ completion of a weak doublet}
\label{sec:model}

We consider the Standard Model (SM) Higgs doublet $H$, a complex SM-singlet scalar $S$ that spontaneously breaks the PQ symmetry, two left-handed Weyl doublets $D,D^c$, a neutral singlet fermion $N$, and a vectorlike colored KSVZ pair $\mathcal Q,\mathcal Q^c$. We define the scalar vacuum expectation values by
\begin{equation}
    \langle H^0\rangle=\frac{v}{\sqrt{2}}\,,\qquad
    \langle S\rangle=\frac{v_S}{\sqrt{2}}\,,
    \label{eq:vevs}
\end{equation}
with $v=246$\,GeV. All SM fields, including $H$, are neutral under $U(1)_{\rm PQ}$. The gauge quantum numbers, PQ charges, and transformation properties under the residual parity are summarized in Table~\ref{tab:charges}. Writing the fermion doublets as
\begin{equation}
    D=\begin{pmatrix}D^0\\
    D^-\end{pmatrix}\,,\qquad D^c=\begin{pmatrix}D^+\\
    D^{c0}\end{pmatrix}\,,
\end{equation}
the interactions relevant for the EW dark sector are
\begin{equation}
    \mathcal L_{\rm dark} = - M_DD\!\cdot\!D^c -\frac{y_S}{2}SNN -y\,NH\!\cdot\!D +\mathrm{h.c.}\,,
    \label{eq:darklag}
\end{equation}
where the dot denotes the antisymmetric $SU(2)_L$ contraction. The charge assignment in Table~\ref{tab:charges} allows all terms in Eq.~\eqref{eq:darklag} while forbidding a bare Majorana mass $NN$ and the Yukawa interaction $NH^\dagger D^c$. After PQ breaking, the singlet acquires the Majorana mass
\begin{equation}
    M_N=\frac{y_Sv_S}{\sqrt{2}}.
    \label{eq:MN}
\end{equation}
The vectorlike doublet mass $M_D$ is PQ invariant and remains independent of the PQ-breaking scale.

The QCD anomaly is generated by the KSVZ fermions through
\begin{equation}
    \mathcal L_{\rm KSVZ} = -y_{\mathcal Q}S\mathcal Q\mathcal Q^c+\mathrm{h.c.},
    \label{eq:KSVZ}
\end{equation}
which gives $M_{\mathcal Q}=y_{\mathcal Q}v_S/\sqrt{2}$ after PQ breaking. The colored fermions are vectorlike under the SM gauge group but chiral under $U(1)_{\rm PQ}$ and generate the QCD anomaly required for the axion solution~\cite{Kim:1979if,Shifman:1979if}. With the down-type gauge quantum numbers chosen in Table~\ref{tab:charges}, PQ-invariant mixing between $\mathcal Q$ and the SM right-handed down-type quarks is allowed and makes the KSVZ fermion unstable. We assume this mixing to be sufficiently small that it does not affect the EW dark sector phenomenology.

In the normalization of Table~\ref{tab:charges}, the color anomaly coefficient has magnitude $|C_{ag}|=2$. Since the field acquiring the PQ-breaking vacuum expectation value has $X_S=2$, the physical domain-wall number is $N_{\rm DW}=1$, and the physical axion decay constant equals $v_S$ in this normalization. The vacuum expectation value of $S$ leaves invariant the transformation $e^{i\pi X_{\rm PQ}}$, defining a residual $\mathbb Z_2$ parity. The fields $N,D,D^c$ are odd under this symmetry, while the SM and KSVZ fields are even. The lightest state in the EW dark sector is consequently stable. This residual parity follows directly from the PQ charge assignment and links the stability of the WIMP component to the same symmetry whose spontaneous breaking produces the QCD axion. In the PQ-broken phase, tree-level elimination of $N$ before expanding $S$ about its vacuum expectation value gives the PQ-invariant interaction $y^2(H\!\cdot\!D)^2/(2y_SS)$. Expanding about $\langle S\rangle=v_S/\sqrt{2}$ and taking $M_N\gg M_D,v$ gives~\cite{Nagata:2014wma, Dedes:2016odh}
\begin{equation}
    \mathcal L_{\rm eff}\supset -M_DD\!\cdot\!D^c + \frac{y^2}{2M_N}(H\!\cdot\!D)^2 +\mathrm{h.c.},
    \label{eq:Leff}
\end{equation}
which is the operator origin of the pseudo-Dirac splitting.

\begin{table}[t]
    \centering
    \renewcommand{\arraystretch}{1.15}
    \begin{tabular}{lccccc}
    \toprule
    Field & $SU(3)_c$ & $SU(2)_L$ & $U(1)_Y$ & $X_{\rm PQ}$ & $\mathbb Z_2$ \\
    \midrule
    $H$              & ${\bf1}$           & ${\bf2}$ & $+1/2$ & $0$  & $+$ \\
    $S$              & ${\bf1}$           & ${\bf1}$ & $0$    & $+2$ & $+$ \\
    $N$              & ${\bf1}$           & ${\bf1}$ & $0$    & $-1$ & $-$ \\
    $D$              & ${\bf1}$           & ${\bf2}$ & $-1/2$ & $+1$ & $-$ \\
    $D^c$            & ${\bf1}$           & ${\bf2}$ & $+1/2$ & $-1$ & $-$ \\
    $\mathcal Q$     & ${\bf3}$           & ${\bf1}$ & $-1/3$ & $0$  & $+$ \\
    $\mathcal Q^c$   & $\overline{\bf3}$  & ${\bf1}$ & $+1/3$ & $-2$ & $+$ \\
    \bottomrule
    \end{tabular}
    \caption{Field content and quantum numbers of the KSVZ realization. All SM fields are neutral under $U(1)_{\rm PQ}$ and even under the residual $\mathbb Z_2$. The fermions $N,D,D^c$ are odd, while the KSVZ fermions are even.}
    \label{tab:charges}
\end{table}
We take $M_D,M_N,y$ real. After PQ and EW breaking, define
\begin{equation}
    m \equiv \frac{yv}{\sqrt{2}}\,.
\end{equation}
The neutral mass matrix in the basis $(N,D^0,D^{c0})$ is
\begin{equation}
    \mathcal M_0=
    \begin{pmatrix}
    M_N&m&0\\
    m&0&M_D\\
    0&M_D&0
    \end{pmatrix}\,.
    \label{eq:M0}
\end{equation}
For $M_N\gg M_D,v$, tree-level matching to Eq.~\eqref{eq:Leff} gives
\begin{equation}
    \mathcal M_{\rm eff} =
    \begin{pmatrix}
    -m^2/M_N&M_D\\
    M_D&0
    \end{pmatrix}\,.
    \label{eq:Meff}
\end{equation}
The two light neutral Majorana mass eigenstates are, up to phase conventions and corrections suppressed by $m/M_N$ and $\delta_\chi/M_D$,
\begin{align}
    \chi_{1L} &\simeq \frac{D^0+D^{c0}}{\sqrt{2}} +\mathcal O\!\left(\frac{m}{M_N}\right)N\,,\\
    \chi_{2L} &\simeq i\,\frac{D^0-D^{c0}}{\sqrt{2}} +\mathcal O\!\left(\frac{m}{M_N}\right)N\,,
    \label{eq:chieigenstates}
\end{align}
with masses
\begin{align}
    m_{\chi_1} &=M_D-\frac{m^2}{2M_N}+\cdots\,,\\
    m_{\chi_2} &=M_D+\frac{m^2}{2M_N}+\cdots\,.
\end{align}
This leads to the mass splitting
\begin{equation}
    \delta_\chi\equiv m_{\chi_2}-m_{\chi_1} = \frac{y^2v^2}{2M_N}\left[1+\mathcal O\!\left(\frac{M_D^2}{M_N^2}\right)\right]\,.
    \label{eq:master}
\end{equation}
In the limit relevant for the phenomenology below, $\chi_1 \equiv \chi$ and $\chi_2$ are the two Majorana components of an almost pure neutral vectorlike doublet, while their singlet admixtures are parametrically of order $m/M_N$ and with the lightest mass
\begin{equation}
    m_\chi =M_D-\frac{\delta_\chi}{2} \simeq M_D\,.
    \label{eq:mchidef}
\end{equation}

For the central kinematic reference benchmark, $M_N=50$\,TeV and $\delta_\chi=324.3$\,keV give $y\simeq2.31\times10^{-2}$ and hence $m=yv/\sqrt{2}\simeq4.03$\,GeV. The singlet admixture of either light state is of order
\begin{equation}
    \frac{m}{\sqrt{2}M_N}\simeq5.7\times10^{-5},
\end{equation}
so the two light states are, to excellent accuracy, pure EW doublet Majorana fermions. Across the representative range $\delta_\chi=304$--$343$\,keV one finds $y\simeq(2.24$--$2.38)\times10^{-2}$. We adopt $M_N=50$\,TeV and $v_S=4.4\times10^{11}$\,GeV as representative choices for the mixed axion--WIMP benchmark, for which $y_S\simeq1.61\times10^{-7}$. The LZ recoil kinematics constrain the combination $y^2/M_N=2\delta_\chi/v^2$, rather than $M_N$ or $v_S$ individually, while the axion abundance fixes a relation between $v_S$ and the initial misalignment angle. The charged Dirac state has tree-level mass $M_D$ and is lifted above the lightest neutral state by an EW radiative correction of order a few tenths of a GeV for the $M_D=297.0$--549.7\,GeV benchmarks, approaching about $0.35$\,GeV in the heavy-doublet limit~\cite{Thomas:1998wy,Ibe:2023dcu}.

The small fermionic Yukawa couplings in this benchmark do not introduce a perturbativity problem between $M_N$ and the PQ scale. In particular, $y\simeq2.3\times10^{-2}$ and $y_S\simeq1.6\times10^{-7}$ give negligible additional contributions to the running of the Higgs quartic~\cite{Abe:2019wku, Wang:2018lhk, Cheng:2019qbd}. We do not require absolute EW vacuum stability up to $v_S$. The scalar potential also allows the portal interaction $\lambda_{HS}|H|^2|S|^2$; for a high PQ-breaking scale an unsuppressed portal introduces the usual EW hierarchy problem of non-supersymmetric axion completions~\cite{Allison:2014hna, Blasi:2024mtc}. Moreover, the portal is not protected by the imposed symmetries and is radiatively generated~\cite{Blasi:2024mtc}. In the present model the $N$--$D$ sector communicates the two scalar sectors through the small Yukawa couplings, with contributions parametrically of order $y^2y_S^2/(16\pi^2)$. We do not attempt to solve the remaining high-scale naturalness problem, which we assume to be controlled by the ultraviolet completion.

\section{DM inelastic scattering in xenon}
\label{sec:dd}

At momentum transfers well below $M_N$, the relevant neutral states are the two Majorana fermions $\chi_1$ and $\chi_2$ in Eq.~\eqref{eq:chieigenstates}. Since they arise predominantly from the two neutral components of the vectorlike EW doublet, the diagonal vector current of the underlying Dirac state becomes, at leading order, an off-diagonal Majorana current,
\begin{equation}
    \mathcal L_Z=i\frac{g}{2c_W} \left[1+\mathcal O\!\left(\frac{m^2}{M_N^2},\frac{\delta_\chi^2}{M_D^2}\right)\right]Z_\mu\bar\chi_2\gamma^\mu\chi_1\,.
    \label{eq:Zcurrent}
\end{equation}
This is the same leading neutral-current structure that governs inelastic scattering of a nearly pure Higgsino~\cite{Fan:2026kxx, Freese:2026sga, Wu:2026nhi, Smirnov:2026aqk}. In the present construction the weak-scale scattering phenomenology is analogous, while the Majorana splitting originates from PQ breaking rather than from heavy EW gauginos. Defining the weak nuclear charge $Q_W=(A-Z)-(1-4s_W^2)Z$ and the normalized weak form factor by $Q_WF_W(q)=(A-Z)F_n(q)-(1-4s_W^2)ZF_p(q)$, the vectorlike EW doublet current in Eq.~\eqref{eq:Zcurrent} gives the zero-momentum nuclear cross section
\begin{equation}
    \sigma_A^0 = \frac{G_F^2\mu_{\chi A}^2}{2\pi}Q_W^2\,,
    \label{eq:sigmaA}
\end{equation}
where $\mu_{\chi A}\equiv m_\chi m_A/(m_\chi+m_A)$. The differential cross section in terms of the nuclear recoil momentum $q=(2m_AE_R)^{1/2}$ is
\begin{equation}
    \frac{\dd\sigma_A}{\dd E_R} = \frac{G_F^2m_A}{4\pi v_{\rm rel}^2}Q_W^2F_W^2(q)\,,
    \label{eq:dsigma}
\end{equation}
where $F_{W,A}(0)=1$. We use the vectorlike doublet normalization of Refs.~\cite{Essig:2007az, Pospelov:2026ewn, Rodd:2026tyn}, which follows directly from the full off-diagonal neutral current of the vectorlike EW doublet. The interaction strength is fixed by the EW neutral current, while the normalization of the event rate depends on the local doublet fraction $\xi_\chi^{\rm loc}$ rather than on an adjustable WIMP--nucleon coupling.

The coherent nuclear response is evaluated with the public \texttt{dmscatter} package~\cite{Anand:2013yka, Gorton:2022eed}, which uses the nuclear response in the proton--neutron basis. Writing $c_p=-(1-4s_W^2)$ and $c_n=1$, we define
\begin{equation}
    \mathcal W_A(q) = \frac{4\pi}{2J_A+1}\sum_{x,x'=p,n}c_x c_{x'}\,W_{M,A}^{xx'}(q)\,,
    \label{eq:weakresponse}
\end{equation}
where $J_A$ is the nuclear spin and $W_{M,A}^{xx'}$ are the isotope-dependent nuclear response functions. Their long-wavelength normalization satisfies
\begin{equation}
    \begin{split}
    W_{M,A}^{pp}(0) &= \frac{2J_A+1}{4\pi}Z^2\,, \qquad W_{M,A}^{nn}(0) = \frac{2J_A+1}{4\pi}N^2\,,\\
    W_{M,A}^{pn}(0) &= W_{M,A}^{np}(0) = \frac{2J_A+1}{4\pi}ZN,    
    \end{split}
    \label{eq:WMzero}
\end{equation}
so that $\mathcal W_A(0)=Q_W^2$. The weak form factor entering Eq.~\eqref{eq:dsigma} is consequently
\begin{equation}
    F_{W,A}^2(q) = \frac{\mathcal W_A(q)}{Q_W^2} = \frac{4\pi}{(2J_A+1)Q_W^2}\sum_{x,x'=p,n}c_x c_{x'}W_{M,A}^{xx'}(q)\,.
    \label{eq:FWdmscatter}
\end{equation}
This construction preserves exactly the zero-momentum normalization in Eq.~\eqref{eq:sigmaA} while replacing the phenomenological Helm suppression by the isotope-dependent nuclear response at finite momentum transfer. We use the \texttt{dmscatter} responses for $^{128}$Xe, $^{129}$Xe, $^{130}$Xe, $^{131}$Xe, $^{132}$Xe, $^{134}$Xe, and $^{136}$Xe isotopes, which account for $99.816\%$ of natural xenon. The residual $^{124}$Xe and $^{126}$Xe components, whose combined natural abundance is $0.184\%$, are described by a Helm response and have a negligible effect on the total rate.

For the endothermic transition $\chi_1 A\to\chi_2 A$, with $\delta_\chi>0$, nonrelativistic energy and momentum conservation require a minimum incident speed for a given recoil energy~\cite{Tucker-Smith:2001myb,Tucker-Smith:2004mxa},
\begin{equation}
    v_{\min}(E_R)=\frac{1}{\sqrt{2m_AE_R}} \left(\frac{m_AE_R}{\mu_{\chi A}}+\delta_\chi\right)\,.
    \label{eq:vmin}
\end{equation}
The function $v_{\min}(E_R)$ reaches its absolute minimum threshold for upscattering $v_{\min}^\star$ at the recoil energy
\begin{equation}
    E_R^\star=\frac{\mu_{\chi A}}{m_A}\delta_\chi,\qquad v_{\min}^\star=\sqrt{\frac{2\delta_\chi}{\mu_{\chi A}}}\,.
    \label{eq:estar}
\end{equation}
At fixed incident speed, the kinematically allowed recoil energies lie between
\begin{equation}
    E_R^\pm(v)=\frac{\mu_{\chi A}^2v^2}{2m_A} \left(1\pm\sqrt{1-\frac{2\delta_\chi}{\mu_{\chi A}v^2}}\right)^2.
    \label{eq:endpoints}
\end{equation}
These relations determine the kinematic support of the recoil spectrum and identify the recoil energy that minimizes $v_{\min}$, but they do not determine the maximum of the spectrum as accepted in the detector. The latter also depends on the finite momentum nuclear response, the Galactic velocity integral, and the LZ efficiency. This distinction is quantified explicitly in Sec.~\ref{sec:rate}. For $E_R\simeq250$\,keV in xenon, the momentum transfer is $q\simeq0.25$\,GeV, where the weak nuclear form factor is already in the diffractive regime and the shell model response differs appreciably from a Helm parametrization. The large value of $v_{\min}^\star$ and the recoil interval in Eq.~\eqref{eq:endpoints} show why sensitivity to splittings of a few hundred keV requires both a heavy target and acceptance at large recoil energy.

LZ provides two-sided intervals for an inelastic $O_1^s$ interaction at $m_\chi=1$\,TeV and an approximate conversion from equal proton/neutron scalar normalization to the weak vector charge,
\begin{equation}
    \sigma_V^{N,\,\rm lim}\simeq\sigma_{\rm SI}^{N,\,\rm lim} \left[\frac{A}{(A-Z)-(1-4s_W^2)Z}\right]^2 \simeq3.2\,\sigma_{\rm SI}^{N,\,\rm lim}\,,
    \label{eq:lzconversion}
\end{equation}
for xenon~\cite{LZ:2026axp}. The corresponding zero-momentum neutron cross section is
\begin{equation}
    \sigma_n^Z=\frac{G_F^2\mu_{\chi n}^2}{2\pi} \simeq 7.43\times10^{-39}\,\mathrm{cm^2}\,,
    \label{eq:sigman}
\end{equation}
for $m_\chi\gg m_n$~\cite{Pospelov:2026ewn}. The indicative upper limit on the local doublet fraction inferred from this zero-momentum rescaling is
\begin{equation}
    \xi_{\chi,\max} = \frac{\sigma_V^{N,\,90\%,\,\rm up}}{\sigma_n^Z}\,.
\end{equation}
The resulting values from the published LZ interval are shown in Table~\ref{tab:lzread}.

\begin{table}[t]
    \centering
    \begin{tabular}{cccc}
    \toprule
    $\delta_\chi$ & $\sigma_{\rm SI}^{90\%,\,\rm up}$ & $\sigma_{\rm weak}^{90\%,\,\rm up}$ & $\xi_{\chi,\max}$\\
    $({\rm\,keV})$ & $(\mathrm{cm^2})$ & $(\mathrm{cm^2})$ & \\
    \midrule
    250 & $1.1\times10^{-42}$ & $3.5\times10^{-42}$ & $4.7\times10^{-4}$\\
    300 & $7.5\times10^{-42}$ & $2.4\times10^{-41}$ & $3.2\times10^{-3}$\\
    330 & $1.0\times10^{-40}$ & $3.2\times10^{-40}$ & $4.3\times10^{-2}$\\
    350 & $6.0\times10^{-40}$ & $1.9\times10^{-39}$ & $0.26$\\
    \bottomrule
    \end{tabular}
    \caption{Upper branches of the two-sided $90\%$ confidence interval read from Fig.~S7 of Ref.~\cite{LZ:2026axp} for $m_\chi=1$\,TeV. The vector-normalized cross sections are obtained using Eq.~\eqref{eq:lzconversion}, and $\xi_{\chi,\max}$ follows by comparison with the fixed EW cross section in Eq.~\eqref{eq:sigman}.}
    \label{tab:lzread}
\end{table}

For a standard thermal doublet at $m_\chi=1$\,TeV, Eq.~\eqref{eq:thermal} gives $\xi_\chi\simeq0.83$. With the vectorlike-doublet normalization in Eq.~\eqref{eq:sigman}, the approximate conversion in Table~\ref{tab:lzread} lies above the published upper interval throughout the tabulated range; even at $\delta_\chi=350$\,keV the corresponding limit is $\xi_{\chi,\max}\simeq0.26$. A thermal TeV-scale doublet would require a splitting still closer to the terrestrial kinematic endpoint. This conclusion differs quantitatively from recasts employing the smaller neutron normalization and illustrates the importance of using the vectorlike EW doublet cross section consistently~\cite{Pospelov:2026ewn}. Since the published LZ interval is given for an $O_1^s$ response at a fixed reference mass, we do not use this approximate conversion to determine a preferred splitting or significance.

The rate calculation presented below instead treats the natural xenon isotopes separately and evaluates their finite momentum weak responses with Eq.~\eqref{eq:FWdmscatter}. This is particularly important in the $E_R\simeq200$--$300$\,keV region, where the relevant momentum transfers probe the second diffractive structure of xenon and small changes in the nuclear response can appreciably modify the event rate. The inelastic kinematics and Galactic velocity integral are evaluated independently of \texttt{dmscatter}; the latter is used only to provide the nuclear response functions. The resulting standard thermal branch, for which $N_{\rm sig}=1$, should be interpreted as a benchmark construction for the fixed EW interaction, with nuclear structure treated explicitly but with residual detector and high velocity tail systematics discussed below.

\section{LZ rate and accepted recoil spectrum}
\label{sec:rate}

To estimate whether the fixed weak interaction can produce an observable yield in the thermal subcomponent regime, we calculate
\begin{equation}
    \frac{\dd R}{\dd E_R}=\frac{\xi_\chi^{\rm loc}\rho_0}{m_\chi} \sum_i\frac{x_i}{\overline m_A}\int_{v>v_{\min}}\dd^3v\,f_{\rm lab}(\bm v,t)\,v\frac{\dd\sigma_{A_i}}{\dd E_R}\,,
    \label{eq:rate}
\end{equation}
where $x_i$ is the natural xenon number fraction and $\overline m_A=\sum_i x_im_{A_i}$. We adopt the Standard Halo Model (SHM) parameters $\rho_0=0.3{\rm\,GeV}\,\mathrm{cm}^{-3}$, $v_0=238\kms$, and $v_{\rm esc}=544\kms$ used by LZ~\cite{LZ:2026axp}. For the rate calculation we approximate the magnitude of the laboratory velocity relative to the Galactic frame by
\begin{equation}
    v_E(t)=254\kms+15\kms\cos[\omega(t-t_0)]\,,
    \label{eq:vEarth}
\end{equation}
with $t_0$ corresponding to the annual maximum in early June. We include the natural xenon isotopes separately, with isotope-dependent weak charges and finite momentum nuclear responses evaluated with \texttt{dmscatter} as described in Sec.~\ref{sec:dd}, except for $^{124}$Xe and $^{126}$Xe, which are computed using Helm form factors. For the benchmarks below we set $\xi_\chi^{\rm loc}=\xi_\chi^{\rm th}$. Defining the annual averaged recoil spectrum by
\begin{equation}
    \overline{\frac{\dd R}{\dd E_R}}
    \equiv
    \frac{1}{T}\int_0^T\dd t\,\frac{\dd R}{\dd E_R}\,,
\end{equation}
the accepted recoil spectrum for an exposure of $\mathcal E=2.84\,{\rm tonne\,yr}$ is
\begin{equation}
    \frac{{\rm d}N_{\rm acc}}{{\rm d}E_R}
    \equiv
    \mathcal E\,\epsilon_{\rm LZ}(E_R)\,
    \overline{\frac{{\rm d}R}{{\rm d}E_R}}\,.
    \label{eq:accepted-spectrum}
\end{equation}
The corresponding signal count is
\begin{equation}
    N_{\rm sig}
    =
    \int_{5.4\,{\rm keV}}^{300\,{\rm keV}}
    {\rm d}E_R\,
    \frac{{\rm d}N_{\rm acc}}{{\rm d}E_R}\,.
    \label{eq:Nsig}
\end{equation}
We define the spectrum peak energy as the nuclear recoil energy at which the annual averaged accepted spectrum is maximal,
\begin{equation}
    E_R^{\rm peak}
    \equiv
    \underset{5.4\,{\rm keV}\leq E_R\leq300\,{\rm keV}}
    {\operatorname{arg\,max}}
    \left[
    \frac{{\rm d}N_{\rm acc}}{{\rm d}E_R}
    \right].
    \label{eq:Epeak}
\end{equation}
Unlike $E_R^\star$, which is fixed purely by the inelastic kinematics through the minimum of $v_{\min}(E_R)$, $E_R^{\rm peak}$ also depends on the velocity distribution, nuclear response, isotope composition, and detector efficiency. Both $E_R$ and $E_R^{\rm peak}$ in Eqs.~\eqref{eq:accepted-spectrum}--\eqref{eq:Epeak} refer to the nuclear recoil energy; the spectrum has not been folded through the resolution or the LZ S1--S2 response. The detector efficiency $\epsilon_{\rm LZ}(E_R)$ is reconstructed from Fig.~S2 of Ref.~\cite{LZ:2026axp}. It remains close to $96\%$ over most of the $14$--$250$\,keV interval and decreases to approximately $50\%$ near $E_R=270$\,keV. We retain the high-energy tail through $300$\,keV, where its contribution becomes negligible.

\begin{figure}[htb]
    \centering
    \includegraphics[width=\linewidth]{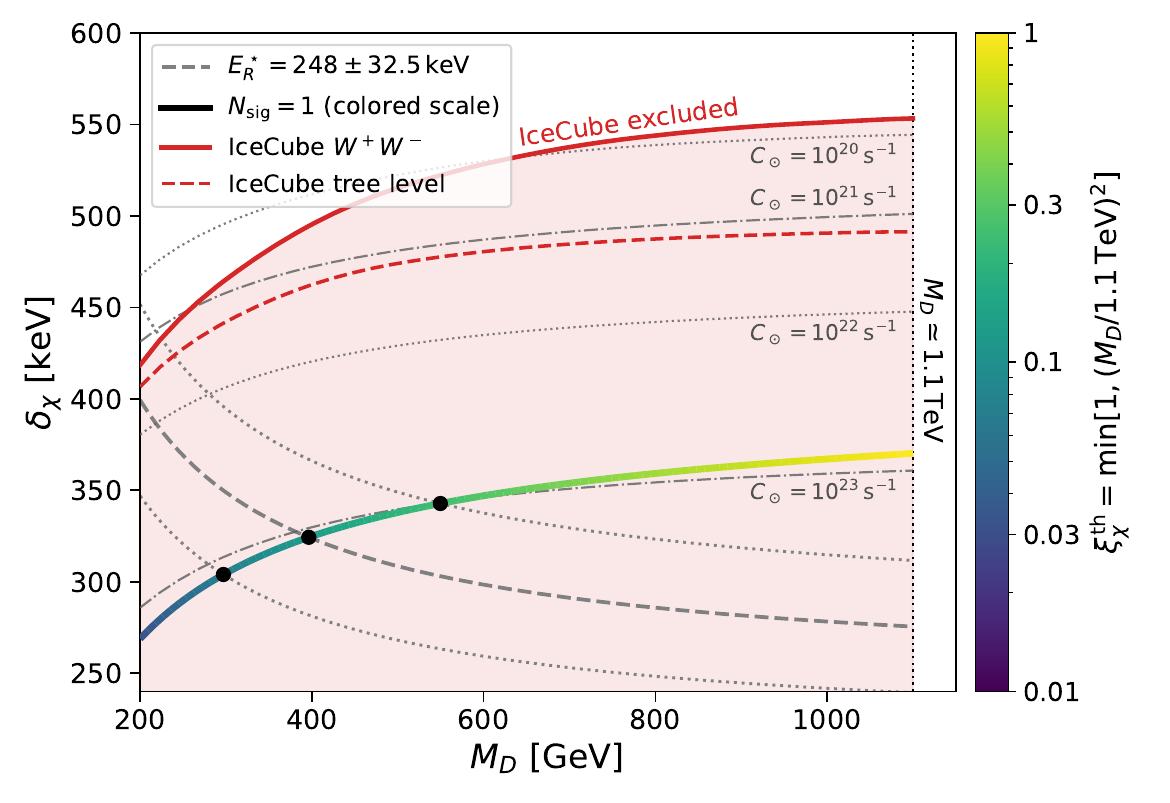}    
    \caption{Standard thermal $N_{\rm sig}=1$ branch and solar neutrino constraints in the $(M_D,\delta_\chi)$ plane. The color scale gives $\xi_\chi^{\rm th}=\min[1,(M_D/1.1\,{\rm TeV})^2]$, with $\xi_\chi^{\rm loc}=\xi_\chi^{\rm th}$. The gray band, bounded by the thin dotted curves, denotes $E_R^\star=248\pm32.5$\,keV for $^{131}$Xe, and the gray dashed curve marks $E_R^\star=248$\,keV. The black points are the kinematic reference benchmarks in Table~\ref{tab:targets}. Gray contours show the solar capture rate $C_\odot$. The solid and dashed red curves denote the IceCube $W^+W^-$ constraints in capture--annihilation equilibrium and using the tree-level stall estimate of Appendix~\ref{app:solar}, respectively. The vertical line marks $M_D=1.1$\,TeV.}
    \label{fig:thermalbranch}
\end{figure}

We keep three kinematic reference benchmarks defined by the intersections of this branch with $E_R^\star=248-\sigma_E$, $248$, and $248+\sigma_E$\,keV for $^{131}$Xe. These values are used only to select representative points spanning the portion of the standard thermal branch most directly associated with the reconstructed recoil scale. The corresponding band in Fig.~\ref{fig:thermalbranch} is retained as a visual guide to the benchmark construction rather than as a likelihood region or a prediction for the maximum of the accepted recoil spectrum. The condition on $E_R^\star$ does not imply that the accepted spectrum peaks at the corresponding recoil energy.

\begin{table}[t]
    \centering
    \begin{tabular}{ccccccc}
    \toprule
    $M_D$ & $\delta_\chi$ & $\xi_\chi^{\rm th}$ & $E_R^\star$
    & $v_{\min}^\star$ & $N_{\rm sig}$ & $y$\\
    $({\rm GeV})$ & $({\rm keV})$ & & $({\rm keV})$
    & $(\kms)$ & & \\
    \midrule
    297.0 & 304.0 & 0.073 & 215.5 & 794.8 & 1.00 & 0.0224\\
    396.4 & 324.3 & 0.130 & 248.0 & 790.5 & 1.00 & 0.0232\\
    549.7 & 342.8 & 0.250 & 280.5 & 785.5 & 1.00 & 0.0238\\
    \bottomrule
    \end{tabular}
    \caption{Kinematic reference benchmarks on the standard thermal branch with $N_{\rm sig}=1$. The three masses correspond to intersections with $E_R^\star=248-\sigma_E$, $248$, and $248+\sigma_E$\,keV for $^{131}$Xe, where $\sigma_E=32.5$\,keV combines the quoted statistical and systematic recoil-energy uncertainties in quadrature. The quantity $E_R^\star$ minimizes $v_{\min}(E_R)$ and should not be identified with the maximum $E_R^{\rm peak}$ of the accepted recoil spectrum. The last column gives $y$ for $M_N=50$\,TeV; $v_S=4.4\times10^{11}$\,GeV corresponds to $y_S\simeq1.61\times10^{-7}$.}
    \label{tab:targets}
\end{table}

\begin{figure}[htb!]
    \centering \includegraphics[width=\linewidth]{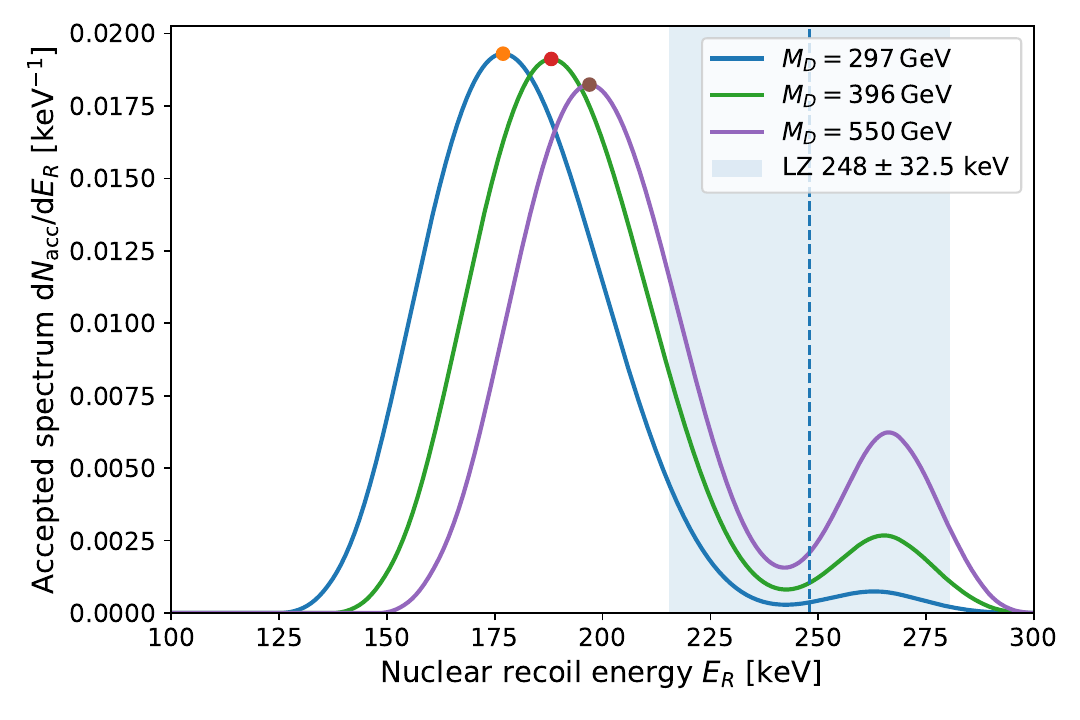}
    \caption{Annual averaged accepted recoil spectra ${\rm d}N_{\rm acc}/{\rm d}E_R$ for the three kinematic reference benchmarks on the standard thermal branch. The vertical dashed line marks $248$\,keV and the shaded interval shows $248\pm32.5$\,keV for comparison with the reconstructed LZ recoil. The spectra have not been folded through the recoil energy response or the LZ S1--S2 likelihood, so the shaded interval is used only as a spectral diagnostic.}
    \label{fig:lzspectra}
\end{figure}

Figure~\ref{fig:thermalbranch} makes explicit both the degeneracy between the doublet abundance and the endothermic suppression and the kinematic prescription used to define the three reference benchmarks. Along the standard thermal branch, increasing $M_D$ raises $\xi_\chi^{\rm th}$, while a larger splitting moves the scattering farther into the high-velocity tail and keeps the integrated accepted LZ count at order unity. The point $(M_D,\delta_\chi)\simeq(396.4\,{\rm GeV},324.3\,{\rm keV})$, with $\xi_\chi^{\rm th}\simeq0.130$, is the intersection with the central kinematic value $E_R^\star=248$\,keV; the two additional black points are defined analogously using $248\pm32.5$\,keV. These points motivate the benchmark selection, while the actual accepted spectral distributions are evaluated independently and shown in Fig.~\ref{fig:lzspectra}.

Maximizing Eq.~\eqref{eq:accepted-spectrum} over the full recoil interval gives $E_R^{\rm peak}\simeq176.9$, $188.1$, and $197.0$\,keV for the three kinematic reference benchmarks, ordered by increasing mass, compared with the corresponding kinematic values $E_R^\star\simeq215.5$, $248.0$, and $280.5$\,keV. The accepted distributions have means $182.1$, $196.3$, and $210.5$\,keV and medians $180.0$, $192.0$, and $203.1$\,keV. Numerical integration of each spectrum reproduces $N_{\rm sig}=1$ to better than $10^{-5}$. The displacement of $E_R^{\rm peak}$ below $E_R^\star$ arises from the interplay of endothermic kinematics with the finite momentum xenon response, the isotope sum, the rapidly varying high-velocity integral, and the detector efficiency. The highest-mass kinematic reference benchmark has the largest accepted fraction in the recoil interval around the observed event, but none of these benchmarks should be interpreted as a spectral best fit. The energy comparison would require folding the recoil spectra through the LZ energy response model and evaluating the likelihood in the experimental observables, which is beyond the scope of this analysis.

The spectra in Fig.~\ref{fig:lzspectra} are annual averages. The instantaneous rate is strongly concentrated near the annual maximum and becomes negligible during part of the year, a characteristic feature of the near-threshold interpretation~\cite{McCabe:2026crm}. The resulting time dependence is highly non-sinusoidal even though Eq.~\eqref{eq:vEarth} uses a single harmonic for the laboratory speed. The occurrence of the candidate on 16 June is consistent with the expected phase, but a single event date contains essentially no modulation information. This is also the regime in which the SHM is least reliable, as small changes in $v_{\rm esc}$ or $v_0$ can shift the splitting by several keV. Simulations further indicate that the Large Magellanic Cloud can contribute to the fastest local DM population, including particles above the nominal Milky Way escape speed~\cite{Besla:2019xbx}. Such a component would increase the rate at fixed $(M_D,\delta_\chi,\xi_\chi^{\rm loc})$ and extend the kinematic reach to larger splittings. The kinematic reference benchmarks in Table~\ref{tab:targets} should not be interpreted as spectral best-fit points, depending on astrophysics and the SHM properties. For the central kinematic reference benchmark, the maximum SHM laboratory speed gives a lower recoil endpoint of about $154$\,keV for $^{131}$Xe. The conventional low-energy WIMP window is kinematically avoided, and the relevant signal lies primarily in the extended high-recoil selection.

\section{The excited state and the LZ event}
\label{sec:decay}

The inelastic collision creates the heavier Majorana state $\chi_2$. Its fate determines whether the event contains only a nuclear recoil or also a delayed electromagnetic (EM) deposit. For the $\mathcal{O}(100{\rm\,keV})$ splittings relevant here, an EW loop involving the charged member of the doublet induces the radiative transition $\chi_2\to\chi_1\gamma$~\cite{Krall:2017xij,Eby:2019mgs,Graham:2024syw}. For a nearly pure pseudo-Dirac EW doublet, the one-loop result can be written as
\begin{equation}
    \Gamma_{\chi_2\to\chi_1\gamma} \simeq C^2(M_D)\frac{e^2g^4}{256\pi^5}\frac{\delta_\chi^3}{M_D^2}\,,
    \label{eq:radiativewidth}
\end{equation}
where $C(M_D)$ is the dimensionless EW loop function. Numerically,
\begin{equation}
    \Gamma_{\chi_2\to\chi_1\gamma} \!\simeq\! 6.4\times10^{-22}{\rm\,GeV}\!\left(\frac{\delta_\chi}{1\,{\rm MeV}}\right)^3\!\!\left(\frac{1\,{\rm TeV}}{M_D}\right)^2\!\!\frac{C^2(M_D)}{C^2(1\,{\rm TeV})},
    \label{eq:radiativewidthnum}
\end{equation}
with $C^2(M_D) \simeq 3.2\left(\frac{M_D}{1\,{\rm TeV}}\right)^{-0.3}$, providing a useful approximation over the range $200{\rm\,GeV}\lesssim M_D\lesssim1{\rm\,TeV}$~\cite{Krall:2017xij}. The singlet admixture of the light states in the present model is only of order $10^{-4}$, while the charged and neutral states have the same leading EW gauge interactions as a pure Higgsino doublet. Corrections to the pure-doublet gauge-loop result are consequently negligible at the accuracy relevant here. In this splitting range the radiative channel also dominates over the tree-level three-body decay.

Using the above approximation, the rest-frame lifetime and laboratory-frame decay length are
\begin{align}
    \tau_{\chi_2} &\simeq 1.61\times10^{-2}\,{\rm s}\left(\frac{M_D}{1\,{\rm TeV}}\right)^{2.3}\left(\frac{400\,{\rm keV}}{\delta_\chi}\right)^3\,,\label{eq:lifetime}\\
    \ell_{\chi_2} &\simeq 6.43\,{\rm km}\left(\frac{v_2}{400\kms}\right)\left(\frac{M_D}{1\,{\rm TeV}}\right)^{2.3}\left(\frac{400\,{\rm keV}}{\delta_\chi}\right)^3\,.\label{eq:decaylength}
\end{align}
The mild deviation from an exact $M_D^2$ scaling originates from the mass dependence of the EW loop function. For the central kinematic reference benchmark, $(M_D,\delta_\chi)=(396.4\,{\rm GeV},324.3\,{\rm keV})$, this gives $\tau_{\chi_2}\simeq3.6\times10^{-3}\,{\rm s}$. At the kinematic threshold the final-state velocity is the center of mass velocity,
\begin{equation}
    v_2^\star\simeq \frac{m_\chi}{m_\chi+m_A}v_{\min}^\star \simeq604\kms\,,
    \label{eq:v2star}
\end{equation}
corresponding to $\ell_{\chi_2}^\star\simeq2.2\,{\rm km}$. For this representative threshold configuration, taking the full $1.5$\,m active-xenon dimension as the available path after the
scatter~\cite{LZ:2019sgr} gives
\begin{equation}
    P_{\rm TPC} \sim \frac{1.5\,{\rm m}}{\ell_{\chi_2}^\star}\simeq6.9\times10^{-4}\,.
    \label{eq:ptpc}
\end{equation}
The variation of the outgoing velocity over the narrow SHM kinematic support changes this estimate only at the $\mathcal O(10\%)$ level. The probability for $\chi_2$ to decay within the active xenon is below $10^{-3}$ and remains small over the scale of a meter, surrounding detector volumes. A correlated EM deposit is consequently not expected for the overwhelming majority of LZ upscattering events, so treating the candidate as a single nuclear recoil is self-consistent.

The photon energy is
\begin{equation}
    \label{eq:egamma}
    E_\gamma=
    \frac{m_{\chi_2}^2-m_{\chi_1}^2}{2m_{\chi_2}}
    =
    \delta_\chi
    \left[1+\mathcal O\!\left(\frac{\delta_\chi}{m_\chi}\right)\right]\,,
\end{equation}
so a decay occurring inside a detector would appear as a narrow EM line near $324.3$\,keV for the central benchmark, with only a small Doppler broadening. The millisecond lifetime also prevents a surviving primordial halo population of $\chi_2$.

Local production can instead occur when $\chi_1$ upscatters from a heavy terrestrial nucleus and the resulting $\chi_2$ traverses a detector before decaying~\cite{Feldstein:2010su,Eby:2019mgs}. The hierarchy $R_{\rm det}\ll\ell_{\chi_2}\ll R_\oplus$ is well satisfied for the kinematic reference benchmarks. Heavy elements such as Pb, Th, and U extend the kinematic reach beyond xenon, while large scintillator detectors provide the volume required to observe the photon~\cite{Graham:2024syw}. Borexino data and the KamLAND, SNO+, and JUNO programs are complementary to LZ~\cite{Eby:2019mgs,Graham:2024syw}. The luminous signal is physically distinct from the LZ nuclear recoil: the collisional rate counts upscattering inside the active xenon, whereas the luminous rate counts terrestrial upscattering followed by decay inside a photon-sensitive volume. Both rates scale linearly with the local fraction $\xi_\chi^{\rm loc}$, but depend differently on target composition, geometry, and the high-velocity distribution. A reanalysis of large scintillator data for a line near $304$--$343$\,keV, including its sidereal-daily modulation, would provide an independent test of the same inelastic splitting~\cite{Eby:2019mgs,Graham:2024syw}.

\section{Cosmological status}
\label{sec:cosmology}

The model contains two cold DM components whose relic abundances originate from distinct cosmological mechanisms. The EW doublet is produced thermally and undergoes conventional freeze-out, whereas the QCD axion is produced nonthermally through vacuum misalignment. The resulting cosmology is a mixed axion--WIMP scenario in which the two components share a microscopic origin in PQ breaking but acquire their relic abundances through independent early-Universe histories.

At freeze-out, the neutral-neutral and charged-neutral mass splittings are both much smaller than $T_f\simeq M_D/25$. EW interactions maintain chemical equilibrium among $\chi_1$, $\chi_2$, and the charged states, so the nearly degenerate multiplet undergoes the standard coannihilation freeze-out~\cite{Gondolo:1990dk, Griest:1990kh, Cirelli:2005uq, Nagata:2014wma},
\begin{equation}
\label{eq:boltzmann}
    \dot n+3Hn=-\langle\sigma_{\rm eff}v\rangle \left(n^2-n_{\rm eq}^2\right)\,,
\end{equation}
where
\begin{equation}
    \label{eq:sigmaeff}
    \langle\sigma_{\rm eff}v\rangle =\sum_{i,j}\langle\sigma_{ij}v\rangle \frac{n_i^{\rm eq}n_j^{\rm eq}}{(n^{\rm eq})^2}.
\end{equation}
Gauge annihilations and coannihilations dominate. Since $\delta_\chi$ and the charged neutral splitting are both much smaller than $T_f$, the standard doublet result already includes efficient coannihilation among $\chi_1$, $\chi_2$, and $\chi^\pm$~\cite{Griest:1990kh,Gondolo:1990dk,Cirelli:2005uq,Nagata:2014wma}. The dimension-five interactions generated by the heavy singlet give negligible corrections to freeze-out because $y^2/M_N=2\delta_\chi/v^2$ is too small to compete with the EW gauge channels. Equation~\eqref{eq:thermal} is not a model identity: it assumes the standard radiation dominated thermal history, entropy conservation after freeze-out, and no additional near-degenerate states. In the present benchmark the singlet lies at $M_N=50$\,TeV and the PQ/KSVZ states are much heavier, so no additional coannihilation partner is present. A precision relic-density calculation should include the physical charged--neutral spectrum and EW Sommerfeld corrections, while non-standard pre-BBN cosmologies, including kination, low-temperature reheating, entropy injection, or late production, can modify the mapping between $M_D$ and $\xi_\chi$ more substantially~\cite{Cirelli:2005uq, Hisano:2006nn, Nagata:2014wma, Visinelli:2015eka}.

For a nearly pure weak doublet and $M_D\lesssim1.1$\,TeV, coannihilation gives the approximate thermal fraction~\cite{Cirelli:2005uq,Nagata:2014wma}
\begin{equation}
    \label{eq:thermal}
    \xi_\chi^{\rm th}\simeq\left(\frac{M_D}{1.1{\rm\,TeV}}\right)^2 \simeq0.073\text{--}0.250\,,   
\end{equation}
where the last step holds for the kinematic reference benchmarks in Table~\ref{tab:targets}. The corresponding relic abundance is $\Omega_\chi h^2\simeq0.12\,\xi_\chi^{\rm th}\simeq0.009$--$0.03$, so the EW component is underabundant and the complementary component is supplied by the QCD axion. For pre-inflationary PQ breaking, vacuum misalignment gives~\cite{Preskill:1982cy,Abbott:1982af,Dine:1982ah,Borsanyi:2016ksw}
\begin{equation}
    \Omega_a h^2\simeq0.12\,\theta_i^2\mathcal F(\theta_i)
    \left(\frac{v_S}{5\times10^{11}{\rm\,GeV}}\right)^{1.165},
    \label{eq:misalignment}
\end{equation}
where $\theta_i$ is the initial misalignment angle and $\mathcal F(\theta_i)$ accounts for anharmonic corrections~\cite{Visinelli:2009zm}. The kinematic reference benchmarks require $\Omega_a h^2\simeq0.09$--$0.11$. For $v_S=4.4\times10^{11}$\,GeV this is obtained for an initial angle of order unity, while $M_N=50$\,TeV corresponds to $y_S\simeq1.61\times10^{-7}$.

The WIMP and axion abundances are not generated by a common freeze-out process: $\Omega_\chi$ is determined predominantly by EW annihilation and coannihilation, whereas $\Omega_a$ is set by vacuum misalignment and controlled by $v_S$ and the initial axion displacement~\cite{Arias:2012az, Marsh:2015xka}. The PQ symmetry nevertheless links the two sectors at the particle-physics level, since the same vacuum expectation value that fixes the axion decay constant generates the heavy singlet mass responsible for the inelastic splitting. The benchmark consequently realizes a genuine mixed axion--WIMP DM scenario in which the two relic components have independent cosmological origins but a common microscopic origin in PQ breaking. For the direct detection calculation we assume that the local WIMP fraction follows its cosmological value, $\xi_\chi^{\rm loc}=\xi_\chi^{\rm th}$. This is a conventional approximation for cold collisionless components rather than an identity, and the two fractions should remain distinct in a global analysis. Since the LZ rate is proportional to $\xi_\chi^{\rm loc}$, departures from this assumption directly rescale the expected yield without modifying the underlying inelastic cross section.

The abundance partition adopted here assumes a standard thermal history for the EW doublet and standard pre-inflationary misalignment for the axion. The colored scale in Fig.~\ref{fig:thermalbranch} should be read as the standard thermal fraction rather than as a fundamental prediction of the PQ construction. Non-standard pre-BBN evolution, including an early matter dominated epoch, modified expansion rate, late entropy injection, or nonthermal production, can alter the WIMP abundance and the onset or dilution of axion oscillations~\cite{Visinelli:2009kt,Arias:2012az,Marsh:2015xka}. Such possibilities would change the quantitative relation between $(M_D,v_S,\theta_i)$ and $\Omega_{\rm DM}$ without altering the PQ origin of the inelastic splitting.

The two components are also subject to distinct additional cosmological consistency conditions. On the axion side, the minimal KSVZ anomaly sector has $N_{\rm DW}=1$. While a post-inflationary PQ transition does not produce a stable string--wall network and is viable, we adopt a pre-inflationary history for the benchmark so that the axion abundance is described by the homogeneous angle $\theta_i$ in Eq.~\eqref{eq:misalignment}. Inflationary fluctuations then generate axion isocurvature perturbations. For small fluctuations,
\begin{equation}
    \label{eq:isocurvature}
    \mathcal P_S\simeq \left(\frac{\Omega_a}{\Omega_{\rm DM}}\right)^2\left(\frac{H_{\rm inf}}{\pi v_S\theta_i}\right)^2\,,
\end{equation}
which is constrained by CMB data~\cite{Planck:2018jri}. For the central kinematic reference benchmark $\Omega_a/\Omega_{\rm DM}\simeq0.87$ and $v_S\simeq4.4\times10^{11}$\,GeV, the isocurvature constraint is not suppressed by a small axion fraction~\cite{Marsh:2015xka, DiLuzio:2020wdo}.

The EW component has a separate cosmological constraint. Once thermal processes that repopulate the excited state become inefficient, $\chi_2$ decays through $\chi_2\to\chi_1\gamma$ on a millisecond timescale, see Eq.~\eqref{eq:lifetime}, while the charged state decays still faster through the soft-pion channel. The fractional energy release in the neutral transition is only $\delta_\chi/m_\chi\sim8\times10^{-7}$. These decays occur well before primordial nucleosynthesis and leave no appreciable cosmological population of excited or charged states. The radiative decay is consequently important for terrestrial experimental signatures but does not generate a late EM energy injection problem for the thermal WIMP component considered here.

\section{Complementarity with other searches}
\label{sec:complementarity}

\subsection{Terrestrial collisional searches}

Complementary experiments probe different aspects of the same inelastic parameter region, but their sensitivity cannot be inferred from exposure alone. Endothermic scattering is favored by heavy nuclei and acceptance at large recoil energy, while for the splittings considered here the event rate is concentrated near the annual maximum of the laboratory speed. LZ combines a multi-tonne xenon target with acceptance extending to recoil energies up to $300$\,keV, and is sensitive to the kinematic reference benchmarks in Table~\ref{tab:targets}. As shown in Fig.~\ref{fig:lzspectra}, however, the accepted spectra are broad and their maxima do not coincide with the kinematic values $E_R^\star$, so comparisons among xenon experiments should use the full accepted recoil distribution rather than a single characteristic recoil energy. Earlier LZ, XENON1T/XENONnT, and PandaX analyses were optimized primarily for standard low-energy WIMP searches~\cite{LZ:2024zvo,XENON:2025vwd,PandaX:2024qfu}, so their published limits cannot in general be compared by a simple rescaling of exposure.

The high-energy events identified in the published LZ, PandaX-4T, and XENONnT WIMP samples provide an additional cross-check~\cite{An:2025bby}. Their recoil energies, selections, and background systematics are not identical to those of the dedicated LZ high-recoil analysis. Extending the xenon experiments to a high recoil window would test whether the $N_{\rm sig}\simeq1$ accepted spectra in Fig.~\ref{fig:lzspectra} predict compatible event populations in data sets with different exposures, efficiencies, and backgrounds. A future multi-tonne xenon observatory would further improve this test through increased exposure and dedicated high-recoil acceptance~\cite{Aalbers:2022dzr}.

Heavier nuclei provide complementary kinematic reach. The PICO-60 CF$_3$I exposure contains iodine and is sensitive to inelastic scattering over a broad recoil-energy range; despite its smaller exposure, the heavy iodine target extends sensitivity toward larger splittings~\cite{PICO:2023uff}. Lead provides still greater kinematic reach. The RES-NOVA PbWO$_4$ cryogenic calorimeter has recently been used to probe large splitting inelastic DM over an extended recoil energy range~\cite{Alloni:2026xdf}. Heavy element paleodetectors provide another complementary strategy, exploiting ancient minerals containing nuclei such as Pb to accumulate inelastic-scattering tracks over geological timescales~\cite{Graham:2026ivn}; their long integration time also makes them sensitive to the time history of the high-velocity DM population, including possible contributions associated with the Large Magellanic Cloud. Tungsten in CRESST provides an even heavier target and can probe regions approaching the kinematic endpoint, while the LZ comparison shows that xenon remains more sensitive over much of the lower splitting region~\cite{LZ:2026axp}. For the present model, these comparisons require the appropriate weak nuclear charge, isotope composition, detector response, and local subcomponent fraction rather than a direct use of published isoscalar limits.

The interpretation of the candidate also remains subject to rare background and detector topology systematics inherent to a single event excess. LZ reconstructs the event $26.4$\,cm above the cathode and well inside the radial fiducial boundary, and finds its S1 partition and S2 pulse shape consistent with a point-like interaction at the reconstructed position~\cite{LZ:2026axp}. The Collaboration finds the event less consistent with multiple-scintillation single-ionization (MSSI) topologies involving charge-insensitive regions. Neutron backgrounds are likewise strongly constrained but not identically zero. These considerations reinforce the importance of an independent test with PandaX-4T or XENONnT rather than altering the particle interpretation developed here.

\subsection{Solar capture and neutrino telescopes}
\label{sec:solar}

DM falling into the solar gravitational potential reaches a local speed
\begin{equation}
    w^2(r)=u^2+v_{\rm esc}^2(r)\,,
    \label{eq:sunvelocity}
\end{equation}
where $u$ is its speed far from the Sun~\cite{Gould:1987ir, Nussinov:2009ft,Blennow:2015hzp,Catena:2018vzc}. Endothermic scattering on a nuclear species $i$ is kinematically allowed only if
\begin{equation}
    \delta_\chi<\frac{1}{2}\mu_{\chi i}w^2\,,
    \label{eq:capturecondition}
\end{equation}
while capture additionally requires the outgoing state to remain gravitationally bound. For the central kinematic reference benchmark, the minimum speeds for scattering on $^{56}$Fe and $^{58}$Ni are about $1.12\times10^3\kms$ and $1.11\times10^3\kms$, respectively. These speeds are reached in the solar interior, so the same splitting that suppresses xenon scattering does not close the solar channel.

We evaluate the solar capture rate across the $(M_D,\delta_\chi)$ plane using the same SHM parameters adopted in the LZ calculation, $\rho_0=0.3{\rm\,GeV\,cm^{-3}}$, $v_0=238\kms$, and $v_{\rm esc}=544\kms$, with a solar frame boost $v_\odot=254\kms$. We set $\xi_\chi^{\rm loc}=\xi_\chi^{\rm th}=\min[1,(M_D/1.1{\rm\,TeV})^2]$ and use the BS2005(AGS,OP) solar profile for the density, composition, and escape-speed structure~\cite{Bahcall:2004pz}. The weak nuclear response is treated at finite momentum transfer using \texttt{dmscatter} where available~\cite{Anand:2013yka,Gorton:2022eed}, supplemented by the \texttt{WIMpy\_NREFT} response tables for the dominant solar targets, including $^{56}$Fe and $^{58}$Ni~\cite{WIMpy-code}; unsupported subleading isotopes are retained with Helm form factors. Details of the capture integral and post-capture evolution are given in Appendix~\ref{app:solar}.

Writing the capture--annihilation evolution as
\begin{equation}
    \Gamma_A=\frac{C_\odot}{2}\tanh^2\!\left[(t_\odot-t_{\rm set})\sqrt{C_\odot C_A}\right],
    \label{eq:solarann}
\end{equation}
the neutrino signal carries one power of $\xi_\chi^{\rm loc}$ once capture and annihilation equilibrate, in contrast with halo annihilation. The solid red curve in Fig.~\ref{fig:thermalbranch} is obtained by comparing the equilibrium $W^+W^-$ partial annihilation rate, $\Gamma_{WW}^{\rm eq} = {\rm Br}_{WW}(M_D)C_\odot/2$, with the mass dependent IceCube $W^+W^-$ annihilation rate limit from the solar DM search over ten years, using the annihilation rate adopted in Refs.~\cite{IceCube:2025fcu,Pospelov:2026ewn}. The dashed red curve applies the same ${\rm Br}_{WW}(M_D)\Gamma_A$ comparison to the tree-level estimate described in Appendix~\ref{app:solar}, including the fixed settling delay. Across the standard thermal $N_{\rm sig}=1$ branch, the capture rate is large enough that the population is close to capture--annihilation equilibrium. The entire branch shown in Fig.~\ref{fig:thermalbranch}, including the kinematic reference benchmarks with $\xi_\chi^{\rm loc}\simeq0.073$--$0.250$, lies below the tree-level estimate contour and is excluded by the adopted IceCube bound.

The comparison uses only the $W^+W^-$ component of the annihilation signal. The accompanying $ZZ$ channel is expected to yield a neutrino signal of comparable strength: the total invisible branching fraction of the $Z$ is about $20\%$, with two neutrinos produced in each $Z\to\nu\bar\nu$ decay, whereas the total leptonic branching fraction of the $W$ is about $30\%$, with one prompt neutrino per leptonic decay. Since a consistent combination would require the channel-dependent neutrino spectra, solar propagation, and detector response, we do not combine the two IceCube templates here. The $W^+W^-$-only comparison reported is conservative with respect to the full EW neutrino signal.

\subsection{Collider and indirect searches}

The charged partner is lifted above the neutral state by the EW radiative correction and, for the resulting $\mathcal{O}(100{\rm\,MeV})$ splitting, decays predominantly through a soft charged pion~\cite{Thomas:1998wy, Ibe:2023dcu}. The collider phenomenology of such compressed EW doublets has been studied using disappearing-track and soft displaced-track signatures~\cite{Mahbubani:2017gjh, Fukuda:2017jmk, Fukuda:2019kbp}. Current ATLAS searches for compressed Higgsino-like spectra exclude chargino masses below $126$\,GeV for splittings between $0.3$ and $2$\,GeV~\cite{ATLAS:2025lhc}, while the CMS search with low-momentum isolated tracks excludes chargino masses up to 185\,GeV for a representative splitting of $0.55$\,GeV and probes splittings between 0.28 and 1.15\,GeV at $m_{\chi^\pm}=100$\,GeV~\cite{CMS:2026ias}. The 297.0--549.7\,GeV benchmarks considered here are not excluded by these searches. A direct application of the simplified-model efficiencies nevertheless requires care, since the physical charged-neutral splitting, lifetime, branching fractions, and production channels must be evaluated for the vectorlike doublet spectrum of the present model.

Halo annihilation proceeds predominantly into EW gauge bosons~\cite{Bertone:2004pz, Hooper:2007qk, Cirelli:2005uq, Nagata:2014wma}. Relative to a single component thermal doublet, the corresponding gamma-ray and cosmic ray fluxes are suppressed by the square of the local abundance fraction, $(\xi_\chi^{\rm loc})^2\simeq0.0053$--$0.063$~\cite{Baum:2016oow}. Solar annihilation has a different scaling: if capture and annihilation reach equilibrium, the annihilation rate is set by the capture rate and carries only one power of $\xi_\chi^{\rm loc}$~\cite{Baum:2016oow, Catena:2018vzc}. Solar neutrino constraints should be treated separately from conventional halo indirect searches.

\section{Discussion and conclusions}
\label{sec:conclusions}

The LZ high-recoil search makes pseudo-Dirac EW DM testable in a regime largely outside conventional WIMP analyses. In the model considered here, a single PQ-breaking scalar sets the QCD axion scale and generates the Majorana mass of a heavy neutral singlet. Integrating out this singlet splits a vectorlike EW doublet, while a minimal KSVZ colored pair supplies the QCD anomaly and the residual PQ parity stabilizes the lighter neutral state. The construction is more constrained than a generic effective inelastic operator: the neutral-current coupling is fixed by the EW quantum numbers, the thermal WIMP abundance follows approximately from gauge annihilation and coannihilation once the doublet mass is specified, and the excited state has a calculable radiative decay.

The LZ interval reported already provides a useful normalization check. For a thermal $1$\,TeV doublet with $\xi_\chi\simeq0.83$, the approximate vector conversion lies above the published upper interval throughout the tabulated range, including at $\delta_\chi=350$\,keV. A doublet near its thermal mass, $M_D\simeq1.1$\,TeV, can constitute essentially the full DM abundance at the level of standard freeze-out, with the LZ rate pushed toward the extreme kinematic endpoint. Such a thermal Higgsino interpretation may be in tension with the absence of events in the higher-energy LZ sideband~\cite{Rodd:2026tyn}. The precise terrestrial splitting depends sensitively on the halo distribution and detector response, while the solar capture analysis of Sec.~\ref{sec:solar} provide an additional and potentially decisive constraint on this full density endpoint.

The sub-TeV portion of the standard thermal branch provides a different realization. Figure~\ref{fig:thermalbranch} follows the standard thermal branch with $N_{\rm sig}=1$ and identifies the kinematic reference benchmarks $M_D=297.0$, $396.4$, and $549.7{\rm\,GeV}$ with $\delta_\chi=304.0$, $324.3$, and $342.8$\,keV, thermal fractions $\xi_\chi\simeq0.073$, $0.130$, and $0.250$, and minimum speeds near $790\kms$. The accepted recoil spectra shown in Fig.~\ref{fig:lzspectra} peak instead at approximately $177$, $188$, and $197$\,keV, with about $6\%$, $16\%$, and $30\%$ of the corresponding accepted rates lying within the numerical interval $248\pm32.5$\,keV. The benchmarks establish rate and kinematic consistency rather than a fit to the reconstructed recoil energy. The excited state decay clarifies the LZ topology rather than complicating it: for the central kinematic reference benchmark the one-loop radiative lifetime is about $3.6\times10^{-3}$\,s and the threshold decay length is about $2.2$\,km, compared with an active-xenon dimension of order $1.5$\,m. The same macroscopic decay length provides a complementary opportunity in large scintillator detectors, where terrestrial upscattering followed by $\chi_2\to\chi_1\gamma$ can generate a narrow electronic-recoil line in the benchmark range 304--343\,keV. A more complete spectral interpretation would require going beyond the kinematic benchmark construction used here. In particular, the predicted recoil spectrum should be folded through the detector response to obtain the distribution in the reconstructed LZ observables, including the recoil-energy resolution, acceptance, and the relevant background model, and the parameters $(M_D,\delta_\chi)$ should then be constrained through an event-level likelihood. Such an analysis would determine the preferred region directly from the measured event rather than from the kinematic quantity $E_R^\star$. This distinction is potentially important because the accepted recoil spectra in Fig.~\ref{fig:lzspectra} peak systematically below the corresponding values of $E_R^\star$. The present points should be regarded as kinematic reference benchmarks on the standard thermal branch, while a likelihood analysis is left for future work.

Independent experimental confirmation is essential, particularly because the present motivation rests on a single event and rare detector background topologies cannot be excluded event by event. Independent analyses in PandaX-4T and XENONnT would test whether a xenon spectrum is reproduced with different backgrounds and exposures, while collider production tests the EW multiplet independently of the Galactic velocity distribution. Solar capture provides a particularly strong complementary test. Using the same SHM parameters as in the LZ calculation, the thermal local fraction at each $M_D$, nuclear responses, and a realistic solar profile~\cite{Bahcall:2004pz}, we find that the standard thermal $N_{\rm sig}=1$ branch is excluded by the IceCube $W^+W^-$ solar neutrino limit, extending the result in Ref.~\cite{Pospelov:2026ewn} for a lower DM abundance since, at the three kinematic reference benchmarks, the tree-level estimate exceeds the IceCube limit by about two to three orders of magnitude. Within the uniform density diagnostic, reducing the signal to the IceCube limit would require an effective radius of about twenty times larger than the corresponding kinematic radius.

The cosmological interpretation varies along the standard thermal branch. For the representative $M_D=297.0$--549.7\,GeV benchmarks, standard freeze-out supplies about 7--25\% of the observed DM density. The QCD axion already present in the KSVZ completion can naturally provide the complementary component: for the representative choice $v_S\simeq4.4\times10^{11}$\,GeV, standard pre-inflationary misalignment yields the required abundance for an initial angle of order unity. The two relic densities arise from independent production mechanisms even though their particle physics origin is linked by the same PQ-breaking vacuum expectation value, leading to a mixed axion--WIMP cosmology. Note that the result obtained extends toward the thermal doublet mass, $M_D\simeq1.1$\,TeV, where $\chi$ can account for all of the DM. Conversely, a lighter doublet could constitute all of the DM only in a non-standard cosmological history that enhances its abundance.

The appropriate claim is narrow and testable. The LZ event motivates a PQ pseudo-Dirac EW doublet whose recoil kinematics, fixed weak normalization, excited state decay, and thermal abundance
are mutually consistent at rate level, while the representative $\mathcal O(100{\rm\,GeV})$ branch realizes mixed axion--WIMP DM. However, under the standard thermal history, the identification $\xi_\chi^{\rm loc}=\xi_\chi^{\rm th}$, the SHM adopted by LZ, and annihilation predominantly into EW gauge bosons, the resulting solar capture rate is incompatible with the IceCube solar neutrino bound along the full standard thermal branch shown in Fig.~\ref{fig:thermalbranch}. A departure from these assumptions can alter this conclusion, but the local WIMP fraction cannot be reduced independently while retaining the same LZ normalization, since both $N_{\rm sig}$ and $C_\odot$ scale linearly with $\xi_\chi^{\rm loc}$. The LZ event provides a concrete target for WIMP DM phenomenology whose interpretation is testable across several fronts: direct detection probes the inelastic recoil and its spectral and temporal structure, solar neutrinos and other indirect searches test the associated annihilation channels, collider experiments probe the EW multiplet, and cosmology constrains the thermal abundance and its embedding in the mixed axion--WIMP scenario.

\begin{acknowledgments}
We thank Michael Zantedeschi for reading a preliminary version of the draft, and Jianglai Liu, Jo\~{a}o Paulo Pinheiro, and Maxim Pospelov for helpful discussions. We also acknowledge Yi-Fu Cai for the hospitality at the University of Science and Technology of China, where this work was carried out. This work was supported by the Istituto Nazionale di Fisica Nucleare (INFN) through the Commissione Scientifica Nazionale 4 (CSN4) Iniziativa Specifica ``Quantum Universe'' (QGSKY), QUAX, and Virgo.
\end{acknowledgments}

\vspace{.3cm}

\section*{DATA AVAILABILITY}

The source code and analysis tools used to generate the results presented in this article are publicly available in Ref.~\cite{visinelli_2026_22759090}.
\vspace{.3cm}

\appendix

\section{Halo and rate conventions}
\label{app:rate}

For completeness, the Galactic-frame distribution used for Table~\ref{tab:targets} is the truncated Maxwellian
\begin{equation}
    f_{\rm gal}(\bm u)= \frac{e^{-u^2/v_0^2}}{N_{\rm esc}\pi^{3/2}v_0^3}\Theta(v_{\rm esc}-u),
\end{equation}
where $z=v_{\rm esc}/v_0$ and
\begin{equation}
    N_{\rm esc}=\operatorname{erf}(z)-\frac{2z}{\sqrt{\pi}}e^{-z^2}.
\end{equation}
The laboratory-frame distribution is $f_{\rm lab}(\bm v,t)=f_{\rm gal}(\bm v+\bm v_E(t))$, with $\bm v_E(t)$ evaluated using the annual velocity prescription described in Sec.~\ref{sec:rate}. Defining the mean inverse speed
\begin{equation}
    \eta(v_{\min},t)= \int_{v>v_{\min}}\frac{\dd^3v}{v}f_{\rm lab}(\bm v,t),
\end{equation}
Eqs.~\eqref{eq:dsigma} and \eqref{eq:rate} give
\begin{equation}
    \frac{\dd R}{\dd E_R} = \frac{\xi_\chi^{\rm loc}\rho_0}{m_\chi}\sum_i\frac{x_i}{\overline m_A}\frac{G_F^2m_{A_i}}{4\pi}Q_{W,i}^2F_{W,i}^2(q_i)\eta\!\left(v_{\min,i},t\right),
\label{eq:appendixrate}
\end{equation}
where $q_i=\sqrt{2m_{A_i}E_R}$, $\overline m_A=\sum_i x_i m_{A_i}$, and $x_i$ denotes the natural number fraction of isotope $i$. The isotope sum contains all naturally occurring xenon isotopes, $A=124,126,128,129,130,131,132,134,136$, with $\sum_i x_i=1$. The count $N_{\rm sig}$ is obtained after applying the recoil-window efficiency described in Sec.~\ref{sec:rate} and averaging the recoil spectrum over a year. Equation~\eqref{eq:appendixrate} makes explicit that the predicted count scales linearly with the assumed local fraction $\xi_\chi^{\rm loc}$. The same Galactic distribution is used for solar capture, with the annual laboratory motion replaced by a solar frame boost of magnitude $v_\odot=254\kms$; no separate halo convention is introduced for the solar calculation.

\section{Solar capture and IceCube limits}
\label{app:solar}

The solar calculation uses the same Galactic distribution as Appendix~\ref{app:rate}. In the solar frame we write the normalized speed distribution as $F_\odot(u)$, obtained by boosting the truncated Maxwellian by $v_\odot=254\kms$. The local number density of the doublet component is
\begin{equation}
    n_\chi^{\rm loc}=\frac{\xi_\chi^{\rm loc}\rho_0}{m_\chi},
\end{equation}
and throughout the standard thermal scan we set
\begin{equation}
    \xi_\chi^{\rm loc}=\xi_\chi^{\rm th}
    =\min\!\left[1,\left(\frac{M_D}{1.1{\rm\,TeV}}\right)^2\right].
    \label{eq:solarxi}
\end{equation}
We use the BS2005(AGS,OP) solar model for the radial density and composition profiles~\cite{Bahcall:2004pz}. The escape speed is reconstructed self-consistently from the enclosed solar mass and satisfies $v_{\rm esc,\odot}(0)\simeq1.38\times10^3\kms$.

For an asymptotic speed $u$, the local speed is $w^2=u^2+v_{\rm esc,\odot}^2(r)$. Defining
\begin{equation}
    w_i'^2=w^2-\frac{2\delta_\chi}{\mu_{\chi i}},
\end{equation}
the scattering channel is open for $w_i'^2>0$, and the recoil endpoints are
\begin{equation}
    E_{R,i}^{\pm}
    =\frac{\mu_{\chi i}^2}{2m_i}\left(w\pm w_i'\right)^2.
    \label{eq:solarERpm}
\end{equation}
A scatter captures the outgoing state only if the total inelastic energy loss is sufficient to remove the positive asymptotic kinetic energy. This gives the additional recoil condition
\begin{equation}
    E_R\ge E_{\rm cap}\equiv\frac{1}{2}m_\chi u^2-\delta_\chi.
\end{equation}
The capture cross section for species $i$ is consequently
\begin{equation}
    \sigma_i^{\rm cap}(r,u)
    =\int_{E_{R,i}^{\rm low}}^{E_{R,i}^{+}}
    \dd E_R\,\frac{\dd\sigma_i}{\dd E_R},
    \qquad
    E_{R,i}^{\rm low}
    =\max\!\left(E_{R,i}^{-},E_{\rm cap},0\right),
    \label{eq:sigmacap}
\end{equation}
when the lower limit lies below $E_{R,i}^{+}$, and vanishes otherwise. The total rate is then evaluated as
\begin{equation}
\begin{split}
    C_\odot =& \sum_i\int_0^{R_\odot}4\pi r^2\dd r\,n_i(r)\nonumber\\
    &\times \int_0^\infty\dd u\,\frac{\xi_\chi^{\rm loc}\rho_0}{m_\chi}\frac{F_\odot(u)}{u}\,w^2(r,u)\,\sigma_i^{\rm cap}(r,u)\,.
    \label{eq:Csun}
\end{split}
\end{equation}
Equation~\eqref{eq:Csun} makes the linear scaling $C_\odot\propto\xi_\chi^{\rm loc}$ explicit. The same off-diagonal weak current and normalization as in Eq.~\eqref{eq:dsigma} are used. At finite momentum transfer, \texttt{dmscatter} responses are used whenever an isotope is available~\cite{Anand:2013yka, Gorton:2022eed}; for the dominant solar nuclei not present in that library we use the $M$-response tabulation distributed with \texttt{WIMpy\_NREFT}~\cite{WIMpy-code}. In particular, $^{56}$Fe and $^{58}$Ni are treated with tabulated nuclear responses rather than a Helm ansatz. Remaining unsupported isotopes are retained with Helm form factors. Along the standard thermal branch, about ninety percent of the total capture rate is carried by nuclei with tabulated finite momentum responses, while the fallback fraction increases toward the largest splittings.

After the first capture event, further inelastic scatters reduce the orbital energy until no solar target can sustain another upscatter. Following Ref.~\cite{Pospelov:2026ewn}, we characterize the terminal tree-level configuration by the $^{238}$U kinematic radius,
\begin{equation}
    v_{\rm esc,\odot}^2(0)-v_{\rm esc,\odot}^2(r_{\rm kin})
    =\frac{2\delta_\chi}{\mu_{\chi U}}.
    \label{eq:rkin}
\end{equation}
An explicit repeated scattering calculation for the kinematic reference benchmarks gives settling times much shorter than the solar age. We impose a fixed delay $t_{\rm set}=0.2$\,Gyr before annihilation; this also covers a stress test in which the adopted uranium abundance is reduced by one order of magnitude. The uranium condition in Eq.~\eqref{eq:rkin} defines a kinematic scale, but by itself does not prove that the complete captured phase-space distribution is
confined within $r_{\rm kin}$. We interpret the following uniform sphere prescription as a tree-level estimate, conditional on confinement within this scale,
\begin{equation}
    C_A^{\rm kin}=\frac{3\langle\sigma v\rangle}{4\pi r_{\rm kin}^3}\,,
\end{equation}
with the benchmark EW annihilation rate
\begin{equation}
    \langle\sigma v\rangle=1.3\times10^{-26}\left(\frac{1.1{\rm\,TeV}}{M_D}\right)^2{\rm cm^3\,s^{-1}}\,.
    \label{eq:sigmavsolar}
\end{equation}
The number of captured particles obeys $\dot N=C_\odot-C_A N^2$, so that
\begin{equation}
    \Gamma_A(t_\odot)=\frac{C_\odot}{2}\tanh^2\!\left[(t_\odot-t_{\rm set})\sqrt{C_\odot C_A}\right]\,.
    \label{eq:GammaAapp}
\end{equation}
The solid IceCube contour in Fig.~\ref{fig:thermalbranch} uses ${\rm Br}_{WW}C_\odot/2$ in the equilibrium limit, while the dashed curve uses ${\rm Br}_{WW}\Gamma_A$ from Eq.~\eqref{eq:GammaAapp} with $C_A=C_A^{\rm kin}$. Since $C_\odot$ already contains the factor $\xi_\chi^{\rm loc}$, the equilibrium neutrino rate scales linearly with the subcomponent fraction; only far from equilibrium does the scaling become quadratic.

For the IceCube comparison we use the mass-dependent $W^+W^-$ annihilation rate limit from the ten-year solar DM search~\cite{IceCube:2025fcu}, in the annihilation rate representation employed in Ref.~\cite{Pospelov:2026ewn}. The EW doublet annihilation rate is shared primarily between the $W^+W^-$ and $ZZ$ final states. At tree-level we write
\begin{equation}
    \label{eq:brww}
    {\rm Br}_{WW}(M_D) = \frac{\mathcal P_W(M_D)}{\mathcal P_W(M_D)+\mathcal P_Z(M_D)/(2c_W^4)},
\end{equation}
where
\begin{equation}
    \mathcal P_V(M_D) = \frac{\left(1-m_V^2/M_D^2\right)^{3/2}}{\left(1-m_V^2/(2M_D^2)\right)^2}\,.
\end{equation}
The quantity constrained by the IceCube limit for a pure $W^+W^-$ template is consequently the partial annihilation rate
\begin{equation}
    \Gamma_{WW} = {\rm Br}_{WW}(M_D)\,\Gamma_A\,.
\end{equation}
The exclusion contour at each $M_D$ is defined by
\begin{equation}
    {\rm Br}_{WW}(M_D)\,\Gamma_A(M_D,\delta_\chi) = \Gamma_{WW}^{\rm IC,lim}(M_D)\,.
    \label{eq:ICcontour}
\end{equation}
To quantify the sensitivity of the IceCube conclusion to this
confinement assumption, we introduce an effective uniform radius
$R_{\rm req}$ through
\begin{equation}
    C_A(R) = \frac{\langle\sigma v\rangle}{(4\pi/3)R^3}\,,
\end{equation}
and define $R_{\rm req}$ by the condition
\begin{equation}
\begin{split}
    &{\rm Br}_{WW}(M_D)\,\frac{C_\odot}{2}\tanh^2\!\left[(t_\odot-t_{\rm set})\sqrt{C_\odot C_A(R_{\rm req})}\right]\\
    &= \Gamma_{WW}^{\rm IC,lim}(M_D)\,.
    \label{eq:Rreq}
\end{split}
\end{equation}
This quantity is a dilution diagnostic rather than an assertion about the true spatial distribution of the captured population. For the three kinematic reference benchmarks in Table~\ref{tab:targets} we find $r_{\rm kin} \simeq (0.182,\;0.176,\;0.170)\,R_\odot$, whereas saturation of the IceCube limit would require $\simeq (3.62,\;3.85,\;3.80)\,R_\odot$. Within the uniform density surrogate, the captured population would have to be diluted over an effective radius about twenty times larger than the tree-level kinematic scale before the predicted $W^+W^-$ neutrino signal is reduced to the IceCube limit. This large separation provides a quantitative robustness check on the solar exclusion without identifying the estimate with a full post-capture phase-space calculation.

For the three kinematic reference benchmarks in Table~\ref{tab:targets}, ${\rm Br}_{WW}\simeq0.545$, $0.543$, and $0.543$, respectively. The remaining annihilation rate is dominated by the accompanying $ZZ$ channel. This channel is not neutrino-poor: ${\rm Br}(Z\to\nu\bar\nu) \simeq0.20$, with two neutrinos produced per invisible $Z$ decay,
whereas the summed leptonic branching fraction of the $W$ is about $0.30$, with one prompt neutrino per leptonic decay. The resulting prompt-neutrino multiplicities from $ZZ$ and $W^+W^-$ are comparable, and the $ZZ$ contribution can only strengthen the solar neutrino constraint at the level relevant here. We nevertheless retain only the $W^+W^-$ component because the IceCube bound employed above is quoted for a pure $W^+W^-$ template; combining the channels consistently would require their propagated neutrino spectra and detector response. The comparison used here is consequently conservative with respect to the full electroweak neutrino signal.

\bibliographystyle{apsrev4-1}
\bibliography{references}

\end{document}